\documentclass[epj]{svjour}
\usepackage{graphics}
\begin{document}
\title{\boldmath $SU(4)$ string tensions from the fat-center-vortices
model}
\titlerunning{$SU(4)$ string tensions}

\author{S.~Deldar\thanks{e-mail: sdeldar@khayam.ut.ac.ir},
S.~Rafibakhsh\thanks{e-mail: shrafi@phymail.ut.ac.ir}}

\institute{Department of Physics, University of Tehran, P.O. Box
14395/547, Tehran 1439955961, Iran}

\date{Received: 30 April 2005 /\\ Published
online:
22 June 2005}

\sloppy\abstract{
The thick- or fat-center-vortices model has been applied to
a calculation of
the potentials
between
static sources of various $SU(4)$ representations.
For intermediate distances, a linear potential is obtained. For
this region
the string tensions agree qualitatively with both flux tube counting
and Casimir
scaling, even though for some representations it favors flux tube
counting more.
In addition, our results confirm
the existence of two different string tensions for non-zero 4-ality
representations at large distances. In this area, zero 4-ality
representations
are screened. In our computations, we have used only the first
non-trivial vortex of $SU(4)$.\\[4mm]
}\maketitle

\section{\boldmath Introduction}
It has almost been proved that QCD is the true theory of
the strong interactions.
For
the high energy regime (asymptotically free region), the
perturbative
formulation shows
impressive agreement between theory and experiment. Lots of efforts
have been
made for the low energy region where a
non-perturbative formulation
is required and
people are interested to see confinement which is one of the
main features of
QCD. It is shown by many numerical measurements in $SU(2)$, $SU(3)$
and $SU(4)$
lattice gauge theory
\cite{Bern82,Deld99} that at intermediate distances a linear
potential between
the quarks for the fundamental and higher representations exists
and the string
tension is representation dependent and roughly proportional
to the eigenvalue
of the quadratic Casimir operator of the representation. This
proportionality
of the potential to the Casimir operator is called ``Casimir
scaling''. On the
other hand some physicists argue for flux counting and describe
the
linear behavior of the potentials at intermediate distances based
on this idea
\cite{Bali00,Shif03} which claims that
the string tension at intermediate distances
is proportional to the number of fundamental flux tubes embedded
into the
representation. Flux tube counting is supposed to coincide with
Casimir scaling
in
the large $N$ limit.

Besides numerical calculations, people have been trying to
introduce phenomenological models to explain
confinement in QCD. One of these models is the center-vortex
theory introduced
in the late 1970's
\cite{Hoof79}.
This model has been developed by Faber {et al.} \cite{Fabe98}
to the
thick-center-vortices model to study the linearity of the potential
for higher
representations
and has been tested for quarks in $SU(2)$.
In this paper, we apply the thick- or fat-center-vortices model
to the sources of
$SU(4)$.
The representations $4$ (fundamental), $6$, $10$, $15$ (adjoint), $20$
and $35$ of
$SU(4)$ are studied. Each representation is shown
by ($n,m$),
where
$n$ and $m$ are the number of original quarks and antiquarks
(in the fundamental
representation) anticipated in producing the representation.
Then since a quark in representation $10$, for example, is constructed
from two
quarks and no antiquark:
\begin{equation}
4\otimes 4= 10 \oplus 6.
\end{equation}
$10$ is shown  by (2,0). Other representations are shown as follows:
$4$: ($1$,$0$), $6$: ($2$,$0$), $15$: ($1$,$1$), $20$: ($3$,$0$)
and
$35$: ($4$,$0$). Our results from the thick-center-vortices model
indicate that
quarks
in all representations of $SU(4)$ are confined at intermediate
distances.
For large distances their behavior is
 based on the 4-ality. Representations
with
zero 4-ality are screened and other potentials will be parallel
to the potential
of either the fundamental representation or representation 6. I recall
that
for the $SU(4)$ gauge group, at large distances two different string
tensions are
expected.
The general behavior of quarks in $SU(4)$ is
in agreement with previous works by Faber {et al.}
\cite{Fabe98} and Deldar \cite{Deld00} for $SU(2)$ and $SU(3)$,
respectively.
At large distances the potential between two sources is large
enough and a pair
of adjoint sources releases from the vacuum. Therefore based
on the $N$-ality
of the representation, we expect to see screening or change of
the slope
of the potential to the slope of the potential in the fundamental
representation. This behavior
has been observed for $SU(2)$, for example, and reported in
\cite{Fabe98},
where at large distances the screening occurs for the adjoint
representation and
the slope of the potential for $j=3/2$ becomes equal to that
of the fundamental
representation. For $SU(3)$,
the representations 3, 6, 8, 10, 15s,
15a and 27 are
studied. Results obtained for
$SU(2)$ and $SU(3)$ at intermediate
distances show
a qualitative agreement with Casimir scaling.  In this paper,
we
show that for $SU(4)$ quarks, one can get two asymptotic string
tensions by
choosing the parameters of
the model appropriately.
As studied for $SU(3)$ \cite{Deld00}, the approximate Casimir
scaling would be
obtained, if one chooses a physical profile for the vortex. For
$SU(4)$,
we use one of the pre-tested profiles which seems physical and
has worked
well for $SU(2)$ and $SU(3)$. There are two different vortices
for $SU(4)$. For
simplicity in the computations, we use only one of them and show
that one can
still get linear potentials for all representations at intermediate
distances.
Comparing the results of the $SU(3)$ and $SU(4)$ gauge groups, we show
that
even though the potentials are proportional to Casimir scaling, a
tendency
to flux tube counting is also observed.

In addition, we take a closer look at the
$SU(3)$ lattice data for
the inter-quark
potentials at intermediate distances. We discuss the possibility
of
string tensions to be proportional to Casimir operators as well
as
the number of fundamental fluxes.

In Sect.\,2, we review very briefly the thick-center-vortices model,
and then the
results of applying this model to $SU(4)$ are given
in Sect.\,3.

\section{\boldmath Calculation of the potentials in
$SU(4)$\newline by the thick-center-vortices
model}

Vortex condensation theory \cite{Hoof79} claims that the QCD
vacuum is filled
with closed magnetic vortices that have the topology of tubes
or surfaces of
finite thickness which carry magnetic flux quantized in the element
of the
center of the gauge group.
In order for the vortex to have a
finite energy per unit length,
the
gauge potential at large transverse
distances must be a pure gauge. However, the gauge transformation
which produces that potential is non-trivial. It is discontinuous
by an element of the gauge center. The non-trivial nature of
the gauge
transformation forces the vortex core to have non-zero energy
and makes
the vortex topologically stable.
A center vortex linked to a Wilson loop, in the fundamental representation
of
$SU(N)$, has the effect of multiplying the Wilson loop by an element
of the
gauge group center, {{i.e.}}
\begin{equation}
W(C) \rightarrow \exp^{\frac{2\pi i n}{N}} W(C)~~~~~~~~n=1,2,...,N-1.
\label{WF}
\end{equation}
The vortex theory states that the area law for the Wilson loop is
due to the
quantum fluctuation in the number of center vortices linking
the loop. The
QCD vacuum does not tolerate a linear potential between adjoint
quarks over
an infinite range. Since adjoint color charges are screened by
gluons, the
force between these sources drops to zero and this is exactly
what happens
in the center vortex theory. However for intermediate distances,
a linear
regime is reported by lattice calculations for quarks of higher
representations. The string tension for quarks of
a higher representation
is obtained if the vortex thickness is quite large: on the order
of the typical
diameters of low-lying hadrons. These vortices are called thick-center
vortices \cite{Fabe98}.
The thick-center-vortices model uses two basic assumptions to
get
confinement for sources of higher representations. The first assumption
describes
that the effect of creating a center vortex piercing the minimal
area of a
Wilson loop may be represented by the insertion of a unitary
matrix at some
point along the loop. This matrix depends on the flux distribution
of the
vortex and the generators of the group in each representation.
The second
assumption says that the probabilities $f$ that loops in the
minimal area are
pierced by vortices and also the random group orientations associated
with
the gauge group elements in each representation are uncorrelated
and should be
averaged.
Then the average Wilson loop is
\begin{equation}
\langle W(C)\rangle
= \prod_{x} \left\{ 1 - \sum^{N-1}_{n=1} f_{n} (1 - {\mathrm {Re}}
{\cal G}_{r}
[\vec{\alpha}^n_{C}(x)])\right\},
\label{WC}
\end{equation}
$x$ is the location of the center of the vortex and $C$
 indicates
the Wilson loop and ${\cal G}_{r}$ is defined as
\begin{equation}
{\cal G}_{r}[\vec{\alpha}] = \frac{1}{d_{r}}
{\mathrm {Tr}} \exp[{\mathrm {i}}\vec{\alpha} . \vec{H}],
\label{gr1}
\end{equation}
where $d_{r}$ is the dimension of the representation and
$\{H_{i},i=1,2,...,N-1\}$  are generators spanning the Cartan
subalgebra. $f$ is the
probability that any given unit is pierced by a vortex.
The parameter $\alpha_{C}(x)$  describes the vortex flux distribution
and
depends on the vortex location; in other words, it depends on
what fraction of
the vortex core is enclosed by the Wilson loop.
Therefore, it  depends on the shape of the loop and the position
of the center
of the vortex in the plane of loop $C$ relative to the perimeter.

On the other hand, the potential may be found by measuring the
Wilson loop and
looking for the area law fall-off at large $T$:
\begin{equation}
W(R,t)\simeq \exp^{-V(R)T},
\label{WV}
\end{equation}
where $R$ is the spatial separation of the quarks, $T$ is the
propagation time,
and $V(R)$
is the gauge field energy associated with the static quark--antiquark
source. For the thick-center-vortices model, $T$ is assumed to be
fixed and very
huge
compared to $R$. Therefore the loop $C$ is just characterized
by the width $R$.
Thus
from (\ref{WC}) and (\ref{WV}) the inter-quark potential
induced by
the vortices is
\begin{equation}
V(R) = \sum_{x}\ln\left\{ 1 - \sum^{N-1}_{n=1} f_{n}
(1 - {\mathrm {Re}} {\cal G}_{r} [\vec{\alpha}^n_{C}(x)])\right\}.
\label{sigmac}
\end{equation}
$V(R)$ depends on
$\alpha_{C}(x)$ (its shape and size) which is determined by the
fraction of the
vortex flux that is enclosed by the Wilson loop.

Vortices of type $n$ and $N-n$ are
the same, except that the magnetic fluxes are in opposite
directions:
\begin{equation}
f_{n}=f_{N-n}~~~~ {\mathrm {and}} ~~~~ {\cal G}_{r} [\vec{\alpha}^n_{C}(x)]=
{\cal G}_{r}^\star [\vec{\alpha}^{N-n}_{C}(x)].
\label{gra2}
\end{equation}
There are three types of vortices in $SU(4)$. Because of
(\ref{gra2}),
$f_{1}=f_{3}$, and
${\mathrm {Re}} {\cal G}_{r} [\vec{\alpha}^1_{C}(x)]=
{\mathrm {Re}} {\cal G}_{r} [\vec{\alpha}^3_{C}(x)]$
and therefore
\begin{eqnarray}
\langle W(C)\rangle
&=&  \prod_{x} \left\{ 1 - 2f_{1} (
1 - {\mathrm {Re}} {\cal G}_{r}[\vec{\alpha}^1_{C}(x)])\nonumber
\right.\\
&& \left. \quad
 \quad
 \quad
- f_{2}
\left(1 - {\mathrm {Re}} {\cal
G}_{r}\left[\vec{\alpha}^2_{C}(x)\right]\right)\right\}.
\label{newWil}
\end{eqnarray}
To make the computation simpler, we assume $f_{2}=0$. This means
that mainly the
first non-trivial center element contributes and in fact it agrees
with other
studies of vortices \cite{Debb02}.
${\cal G}_{r}$ for vortex number 1 is
\begin{equation}
\label{gr2}
{\cal G}_{r}[\vec{\alpha}] = \frac{1}{d_{r}} {\mathrm {Tr}}
 \exp[{\mathrm {i}}(\alpha^1_{1}
H_{1}+
\alpha^1_{2} H_{2}+\alpha^1_{3} H_{3})].
\end{equation}
The upper index of $\alpha$ shows the vortex profile for vortex
number $1$ which is associated with $f_{1}$. If one wants to
include vortex
number $2$ in the computations, one may choose another vortex
profile
with upper index 2. Vortex number $2$ is associated with $f_{2}$.
The diagonal elements of the Cartan subalgebra in $SU(4)$ are
\begin{eqnarray}
H_{1} & : &  \frac{1}{2\sqrt{6}}(1,1,1,-3) ,\nonumber \\
H_{2} & : &  \frac{1}{2\sqrt{3}}(1,1,-2,0) ,\nonumber \\
H_{3} & : &  \frac{1}{2}(1,-1,0,0).
\end{eqnarray}
Only $H_{1}$ is a $4*4$ matrix with all
diagonal elements
non-zero,
and the
other two come from the $SU(2)$ and $SU(3)$ gauge groups. Because
of this fact,
one may use only $H_{1}$ and $\alpha^1_{1}$ in the
calculations and
thus
(\ref{gr2}) reduces to
\begin{equation}
{\cal G}_{r}[\vec{\alpha}] = \frac{1}{d_{r}}
{\mathrm {Tr}} \exp[{\mathrm {i}} \alpha^1_{1}
H_{1}].
\label{gr}
\end{equation}
The real part 
of the trace of ${\cal G}_{r}[\vec{\alpha}]$ in (\ref{gr}),
with appropriate normalization factor of $\alpha$, contains
the real part 
of the trace of
(\ref{gr2}). So again, for simplicity we use
(\ref{gr}).
Calculating the normalization factor for $\alpha$'s is discussed
later.

The appropriate flux distribution,
$\alpha(x)$,
can be chosen
such that
a well behaved potential is obtained.
This means that with a good choice, one can see
the linear term of the potential for all representations.
In general any physical axially symmetric density distribution
for the vortex
leads to potentials which are acceptable in QCD.
A variety of different fluxes is introduced in
\cite{Deld00}. Here we choose $\alpha_{c}(x)$ to
be
\begin{equation}
\alpha_{R}(x)= (\frac{\sqrt{6}\pi}{2}) [1-\tanh(ay(x)+ \frac{b}{R})],
\label{alphar}
\end{equation}
$a$, $b$ and  $f$ are constants and are free parameters of the
model. They must
be chosen such that a linear potential at intermediate distances
as well as
proportionality with Casimir scaling are observed.
In this paper $a=0.05$, $b=4$, $f_{1}=0.1$ and
\begin{equation}
y(x)= \left \{
\begin{array}{ll} x-R ~~ \mbox{for $|R-x| \leq |x|$ } ,\\
-x ~~ \mbox{for $|R-x| > |x|$ } ,
\end{array}
\right.
\end{equation}
The normalization factor for $\alpha$ is obtained based on the assumptions
of the
model. Equation\,(\ref{WF}) shows that each time that a vortex links
to a
fundamental Wilson loop, the loop is multiplied by the factor
$\exp^{\frac{2\pi i n}{N}}$. Since we are using only one vortex and $N=4$, we have:
\begin{equation}
W(C) \rightarrow \exp^{\frac{\pi i }{2}}W(C) .
\end{equation}
Thus, based on the
thick-center-vortices model, for higher representations,
every
time that the minimal surface is pierced
by a center vortex, a center element $\exp({\frac{\pi i }{2}})$ should
be inserted somewhere along the loop:
\begin{equation}
W(C)=Tr[UU...U]\rightarrow Tr[UU...\exp[{\frac{\pi i }{2}}]...U].
\end{equation}
Using this fact and the point that we have used only one vortex
and only
$H_{1}$ in our computations, the normalization factor is obtained
from
$\exp^{\frac{\pi  {i} }{2}}=\exp^{
{i} \alpha H_{1} }$ which happens
if the vortex
core is entirely contained within the Wilson loop. Therefore
$\alpha$ should
satisfy the following conditions.
\\
\noindent
(1) Vortices which pierce the plane far outside the loop do not
affect the loop.
In other words, for fixed $R$, as $x\rightarrow \infty$,
$\alpha \rightarrow 0$.
\\
\noindent
(2) If the vortex core is entirely contained within the loop,
then
$\alpha=\sqrt{6}\pi$.
\\
\noindent
(3) As $R\rightarrow 0$ then $\alpha \rightarrow 0$.

Equation\,(\ref{alphar}) with its normalization factor satisfies
the above
conditions.
Changing the parameters $a$ and $b$ by any factor $F$
and setting $a \rightarrow \frac{a}{F}$ and $b\rightarrow bF$ only
changes the
scale
of the potential and
the physics remains the same. However not any
arbitrary value
for $a$ and $b$ leads to a linear potential.

\begin{figure}
\begin{center}
\vspace{70pt}
\resizebox{0.47\textwidth}{!}{
\includegraphics{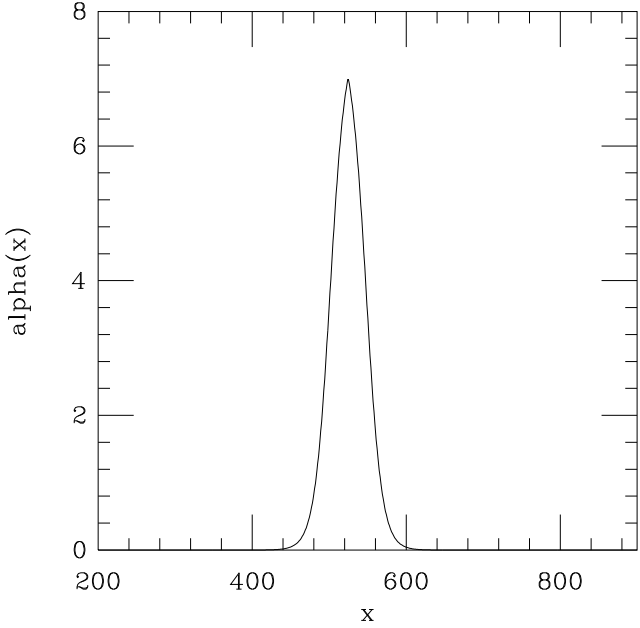}}
\vspace{-20pt}
\caption{\label{fig:flux}
Flux distribution of (\ref{alphar}). For this plot $R=50$ and
the
Wilson loop may entirely overlap the vortex core}
\end{center}
\end{figure}

\begin{figure}
\begin{center}
\vspace{70pt}
\resizebox{0.47\textwidth}{!}{
\includegraphics{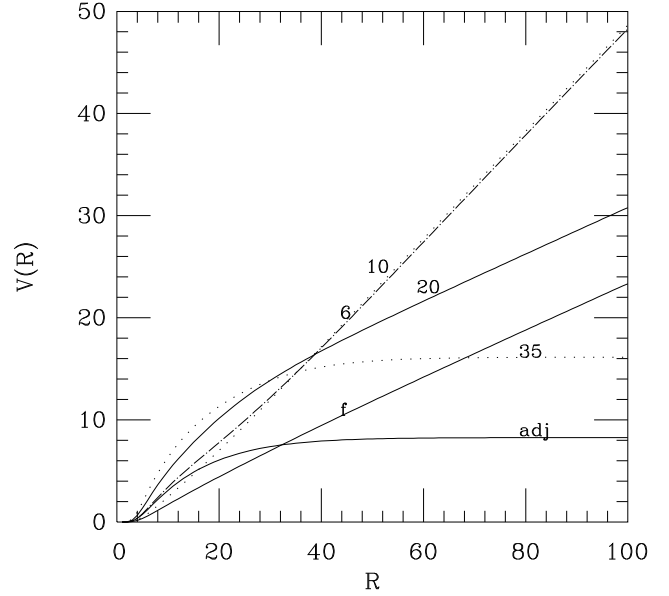}}
\vspace{-20pt}
\caption[]{\label{fig:fig1}
Static sources potential for the range of $R \in [1,100]$. At
intermediate distances, potentials are linear. At large distances,
representations 15 (adjoint) and 35 are screened; representation 20
gets the
same slope as the fundamental representation and representation
10 is parallel
to representation 6}
\end{center}
\end{figure}

Figure\,\ref{fig:flux} shows this flux distribution versus $x$,
the
location of the center of the vortex. For this plot $R=50$. Thus
the size
of the loop is such that the vortex may overlap completely the
Wilson loop.
Another example of a physical flux distribution is the one which
is zero
everywhere except on the boundary of the vortex \cite{Deld00}.
A density proportional to a delta function which is zero everywhere
except
at the two points where the vortex first enters and exits the
Wilson loop is among the non-axially symmetric distributions.
Using this flux,
one loses Casimir scaling proportionality.

\section{\boldmath Results and discussion}

Figure\,\ref{fig:fig1} shows the potentials for various representations
versus $R$ in
the range
$R \in [1,100]$. For all representations, there is a
region where the potentials are linear (about $R \in [5,15]$).
In general,
at large distances, screening for $15$ (adjoint) and $35$ should
occur.
4-ality which is $(n,m)$ mod $4$, is zero for representations
$15$: $(1,1)$
and $35$: $(4,0)$. Screening is expected for zero 4-ality representations.
In
fact,
at large distances, the potential between two sources is large
enough to
create a pair of quarks in the adjoint representation ($15$)
from the vacuum.
Then the original static sources combine with the $15$s and for
zero 4-ality
representations a $15$ component is created and the potential
will
be screened. For non-zero 4-ality representations, two possible
scenarios may
occur: for example, for representation 20, a fundamental component
will be
created and therefore the slope of the potential changes to that
of the
fundamental representation; on the other hand, for $6$ and $10$
dimensional
diquark representations, a quark in representation $6$ is created
and the
slope of the potential in representation $10$ changes to the
slope of
the potential in the representation $6$.
We have
\begin{eqnarray}
15 \otimes 6  & = &  64 \oplus  10 \oplus \bar{10} \oplus 6,
\\
15 \otimes 10  & = &  70 \oplus  64 \oplus \bar{10} \oplus 6.
\end{eqnarray}
This is why there are two different string tensions, one for
the $4$
and another for the $6$ dimensional representation. For
the $SU(N)$ gauge
group
with $N>3$, there is more than one asymptotic string tension.
In $SU(4)$, potentials of the $6$ and $10$ dimensional representations
do not parallel the fundamental representation potential
and the slope of the potential of representation $10$ changes
to that of
representation $6$ \cite{Ohta99}.
Using only vortex number 1 in our calculations, $f_{2}=0$, we
have gotten two
different string tensions for the $SU(4)$ sources.

\begin{table}
\caption[]{\label{tab:tab1}
This table shows Casimir numbers ratios, number of flux tubes
and
string tensions ratios for the
$SU(4)$ gauge group. A qualitative agreement
of string
tensions with both Casimir scaling and flux tube counting is
observed}
\tabcolsep=2pt
\begin{center}
\begin{tabular}{@{}llccccc@{}}
\hline
Repn.&  $4$ (fund.) & $6$ & 15a
 & $10$&  $20$ & $35$ \\
\hline
$(n,m)$    &  $(1,0)$ & $(2,0)$ & $(1,1)$ &  $(2,0)$ & $(3,0)$
& $(4,0)$ \\ \\
$c_{r}/c_{f}$   &   $1$ &   $1.33$ &  $2.13$ &  $2.4$ &   $4.2$
&  $6.4$  \\
\\fund. fluxes &   $1$ &   $2$  &  $2$ &  $2$     &  $3$
&  $4$    \\
$k_{r}/k_{f}$   &    $1$ &  $1.51$ &    $1.56$ & $1.76$  &  $2.31$
&  $2.66$
\\
\hline
\end{tabular}
\end{center}
\end{table}

\begin{table}
\tabcolsep=2pt
\caption[]{\label{tab:tab2}The same as Table\,\ref{tab:tab1}
but for the $SU(3)$
gauge group}
\begin{center}
\begin{tabular}{@{}llcccccc@{}}
\hline
Repn.&  3 (fund.) & $8$ & $6$ & 15a&  $10$ & $27$ & 15s
\\
\hline
$(n,m)$    &  $(1,0)$ & $(1,1)$ & $(2,0)$ &  $(2,1)$ & $(3,0)$
& $(2,2)$ &
$(4,0)$ \\
$c_{r}/c_{f}$   &   $1$ &   $2.25$ &  $2.5$ &  $4$ &   $4.5$
&  $6.$ &  $7$ \\
fund. fluxes &   $1$ &   $2$  &  $2$ &  $3$     &  $3$   &  $4$
& $4$   \\
$k_{r}/k_{f}$   &    $1$ &  $2.02$ &    $2.21$ & $3.1$  &  $3.4$
&  $3.8$ &
$5.6$ \\
\hline
\end{tabular}
\end{center}
\end{table}

\begin{figure}
\begin{center}
\vspace{70pt}
\resizebox{0.47\textwidth}{!}{
\includegraphics{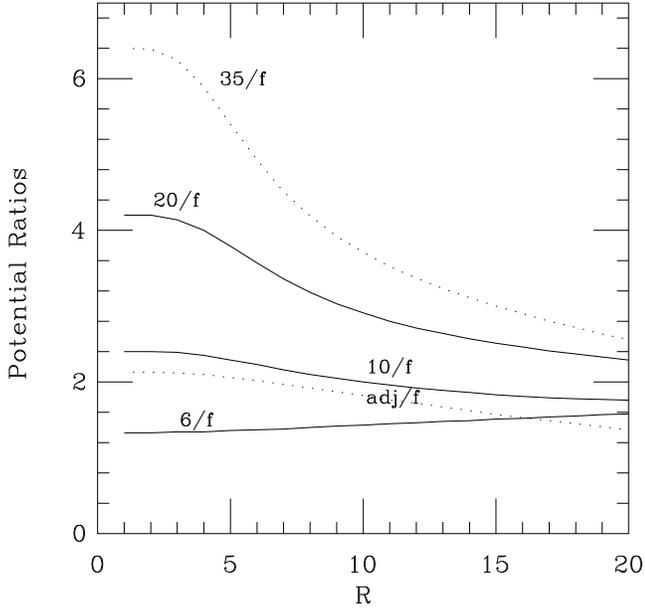}}
\vspace{-20pt}
\caption[]{\label{fig:fig2}Potential ratio of each representation
to the
fundamental representation. These ratios start up roughly at
the ratios of
the corresponding Casimirs, which are $1.3$, $2.4$, $2.13$, $4.2$
and $6.4$
for representations $6$, $10$, $15$, $20$ and $35$, respectively,
but in the
most linear part of the potential (Fig.\,\ref{fig:fig3}),
the ratios are also in agreement with the flux tube counting
especially for the
higher
representations. The slope of the potentials will become constant
after the confinement regime is finished. It happens at about
$R=30$ with
the given parameters $(a,b,f)$ we used in the model}
\label{ratio}
\end{center}
\end{figure}

\begin{figure}
\begin{center}
\vspace{70pt}
\resizebox{0.47\textwidth}{!}{
\includegraphics{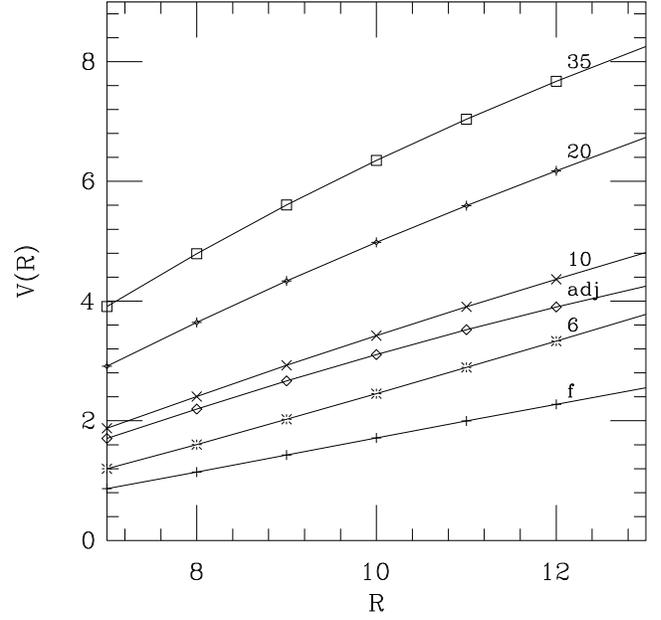}}
\vspace{-20pt}
\caption[]{\label{fig:fig3}The most linear part of the potentials.
The slope of
each potential is
obtained for the range $R \in [7,12] $. They are .282, .427,
.439, 498, .652
and .752 for representations fundamental, 6, adjoint, 10, 20
and 35,
respectively. The ratios of string tensions (slopes of the linear
part of
potentials) are reported in Table\,\ref{tab:tab1}}
\end{center}
\end{figure}

Figure\,\ref{fig:fig2} plots the ratios of the potential
of each representation to that of the fundamental one.
Although these ratios
start up roughly at the ratios of
the corresponding Casimirs, which are
$1.3$,
$2.4$, $2.13$, $4.2$ and $6.4$ for the representations $6$, $10$,
$15$, $20$ and
$35$, respectively, but in the most linear part of the potential
(Fig.\,\ref{fig:fig3}),
the ratios of some representations are closer to the flux tube
counting.
Table\,\ref{tab:tab1} shows the representations we have studied
with the number
of
original quarks and antiquarks, $(n,m)$,
 anticipated in each representation.
In the third row, the ratio of Casimir number of each representation
to that of
the
fundamental one is indicated. The number of fundamental
fluxes existing
in each
representation
is shown in the fourth row. The last row indicates the ratio
of the string
tension
of each representation to the fundamental one, calculated from
this work. These string tensions are obtained from the range
of $R \in [5,12]$ which is the most linear part of the potentials
in our
calculations. The agreement with both Casimir scaling and
flux counting is qualitative, but it favors flux tube counting
more.
Comparing Tables\,\ref{tab:tab1} and \ref{tab:tab2}, one can
see that the tendency
to the flux tube counting increases by increasing the number
of gauge groups,
especially for the representations $35$ of $SU(4)$ and 15s of $SU(3)$.

\begin{figure}
\begin{center}
\vspace{70pt}
\resizebox{0.47\textwidth}{!}{
\includegraphics{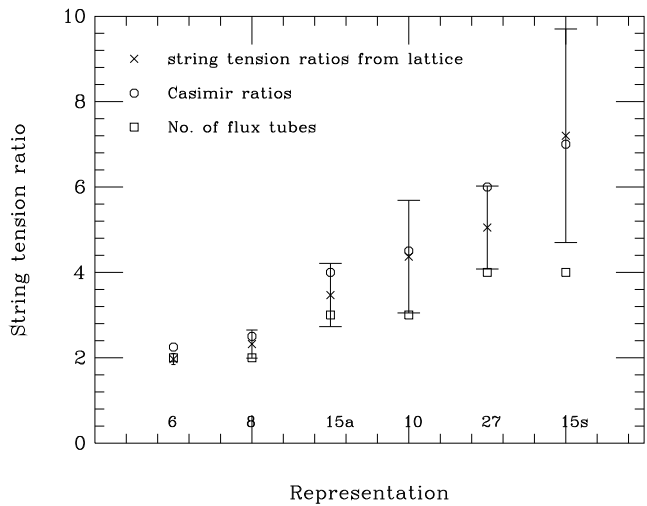}}
\vspace{-20pt}
\caption[]{\label{fig:fig5}Ratios of string tensions of $SU(3)$
quarks of
different representations to the string tension of quarks in
the fundamental
representation are plotted. Considering the lattice data errors,
good
agreement with both
Casimir scaling and flux tube counting is observed
}
\end{center}
\end{figure}

Looking closer at Fig.\,\ref{fig:fig2}, it seems that there
are four regions
for the potentials. For the first area, basically for
$R<5$, the potentials
are proportional to Casimir scaling; especially for very small
$R$.
The potential ratios start out at the ratios of Casimirs
which are larger than the number of fundamental fluxes, but they
get close to the number of fluxes embedded in each representation
at
about $R=10$ for the representations 35, 20 and 10 and at about $R=20$
for
representations 15 (adjoint) and 6.
Based on the assumptions of the thick-center-vortices model,
vortices are
uncorrelated and also the orientation of the ``vortex insertion''
in color
space is chosen at random. Thus, it might be possible that these
assumptions
lead effectively to the interaction of flux tubes. Then one may
explain the
behavior of the potentials at different distances by the interaction
between
fundamental strings as follows:
When the distance between sources is very small, the fundamental
fluxes overlap
and the string tension is larger than the number of fluxes times
the fundamental string tension. This is because of the repulsion
between
the fluxes \cite{Shif03}. This may explain the behavior of the potential
for
distances less than
$5$ in Fig.\,\ref{fig:fig2}. On the other hand
as the distance increases, the fluxes tend to attract each other
and therefore
a negative energy is added to the binding energy of fluxes and
this
makes the string
tension smaller (for $10<R<20$) \cite{Shif03,Fabe93}.
In general if there is no interaction between
the fluxes, the string
tension of
the representation must be $K\sigma_{f}$ where $K$ is the number
of fundamental
fluxes and $\sigma_{f}$ is the fundamental string tension.
Screening or change of the slope of
the potentials starts from about
$R=30$.

Casimir scaling obtained from $SU(3)$ lattice calculations by
Bali
\cite{Bali00} has been verified for
distances less than $1$\,fm and the ratios are reported
to be bigger than
the number
of fluxes.
It seems that the thermal distance between strings in the lattice results
is not
large enough and an overlap between
the strings leads to a repulsion
force
and therefore increasing of the string tension. This coincides
with the $R<5$ of
Fig.\,2. It should be noticed that the scales of $R$ and $V(R)$
are
arbitrary (adjustable) in the thick-center-vortices model.
On the other hand, the results reported by Deldar \cite{Deld00}
are obtained
for larger distances (up to $2.5$\,fm).
Figure\,\ref{fig:fig5} is plotted based on the data of
\cite{Deld00}.
Cross signs show the ratios of the
string tensions of $SU(3)$ sources
of various
representations to that of the fundamental representation. Circles
indicate
Casimir ratios, and diamonds show the number of fundamental strings
in each
representation.
As claimed by the lattice people, the ratios of
the string tensions are proportional to the
Casimir operators. On the
other hand,
the plot shows that there is a rough agreement with the number
of
fundamental tubes as well.  One may conclude that the errors
of the
lattice data are still too large to discriminate between the
two hypotheses,
Casimir scaling and flux counting.

\section{\boldmath Conclusion}

Using thick-center-vortices model, we have shown that for all
representations
at intermediate distances there is a region where quarks are
confined and
the potentials are linear. For this linear regime, our results show
evidence of
proportionality of the string tensions with both Casimir scaling
and
flux counting even though the agreement is slightly better for
the
flux tube counting.
At large
distances, zero 4-ality representations ($15$, $35$) are screened,
the potential of the representation $20$ has become parallel to that
of the fundamental
representation and
the potentials of the diquark representations ($6$ and $10$) get the
same slope.
In our computations, we have used only the first non-trivial
vortex of $SU(4)$.
Although we have not used vortex number $2$ in our calculations,
our results at intermediate distances agree
with lattice results and
the potentials
are linear and qualitatively in agreement with Casimir scaling
at intermediate
distances.

We are doing new computations with $f_{2}$ not necessarily zero
and study the
effect of vortex number 2 and also the different core sizes
of the vortices in
the linear region of the potential. Especially,
we are interested in studying the effect of different parameters
of the model
on removing the concavity of the potentials which has been observed
for $SU(2)$
and $SU(3)$ sources, as well.

\section{\boldmath Acknowledgement}

We would like to thank Dr. S. Olejnik and Dr. M. Faber for their
valuable
comments and patience in answering our questions.
We are grateful to Dr. C. Bernard for his help, especially his
suggestion
of studying $SU(4)$ potentials.
We also thank the research council of the University of Tehran for
support
of this work.

\end{document}